\documentclass[a4paper,10pt]{article}

\input epsf.tex

\newcommand{\be}{\begin{equation}}
\newcommand{\ee}{\end{equation}}
\newcommand{\ba}{\begin{eqnarray}}
\newcommand{\ea}{\end{eqnarray}}
\newcommand{\ban}{\begin{eqnarray*}}
\newcommand{\ean}{\end{eqnarray*}}

\newcommand{\sandwich}[3]{\mbox{$ \langle #1 | #2 | #3 \rangle $}}
\newcommand{\ket}[1]{\mbox{$ | #1 \rangle $}}
\newcommand{\bra}[1]{\mbox{$ \langle #1 | $}}

\newcommand{\si}{\sigma}
\newcommand{\demi}{\frac{1}{2}}

\newcommand{\one}{\leavevmode\hbox{\small1\normalsize\kern-.33em1}}

\newcommand{\Tr}{\mbox{Tr}}

\newcommand{\moy}[1]{\langle #1 \rangle}

\begin{document}

\title{\Large \sc{Quantum Key Distribution between N partners: optimal eavesdropping and Bell's inequalities}}

{\normalsize{\author{ Valerio Scarani\thanks{corresponding
author. Tel. +41 22 7026883; fax +41 22 7810980; e-mail:
valerio.scarani@physics.unige.ch}, Nicolas Gisin \\
Group of Applied Physics, University of Geneva\\
20, rue de l'Ecole-de-M\'edecine, CH-1211 Geneva, Switzerland\\
 }}}
\maketitle

\begin{abstract}
Quantum secret-sharing protocols involving N partners (NQSS) are
key distribution protocols in which Alice encodes her key into
$N-1$ qubits, in such a way that all the other partners must
cooperate in order to retrieve the key.
On these protocols, several eavesdropping scenarios are possible:
some partners may want to reconstruct the key without the help of
the other ones, and consequently collaborate with an Eve that
eavesdrops on the other partners' channels. For each of these scenarios, we
give the optimal individual attack that the Eve can
perform. In case of such an optimal attack, the
authorized partners have a higher information on the key than the
unauthorized ones if and only if they can violate a Bell's
inequality.
\end{abstract}

\section{Introduction}

In the rapidly growing field of quantum
information, the first protocol that has
almost reached the level of application is quantum cryptography
\cite{revue}, a beautiful solution to the important problem of
secure communication. The authorized
partners Alice and Bob can establish an absolutely secure
communication provided that they share a common sequence of bits
(the {\em key}), unknown to anybody else: this is the very
principle of the so-called secret-key cryptographic schemes. In 1984, Bennett and Brassard \cite{bb84} proposed a way of
distributing the key in a physically secure way by using quantum
physics: their protocol bears the acronym BB84, and was the first protocol of quantum cryptography ---
from now on, we shall use the more precise name of {\em quantum key distribution}
(QKD). In the intuition of Bennett and Brassard, security is provided by the
well-known feature of quantum mechanics: "measurement perturbs the
system"; or, under a different viewpoint which is equivalent, by
the no-cloning theorem. In 1991, Ekert \cite{ekert} proposed a QKD protocol
that uses entangled particles, and stated that the
violation of Bell's inequality might be the physical principle
that ensures security. This view was challenged by Bennett,
Brassard and Mermin \cite{bbm}, who showed that Ekert's protocol is actually
equivalent to the BB84 protocol, that involves single particles.
A link between security of QKD and Bell's inequalities was
nevertheless noticed in further studies \cite{bruno,fuchs}.

The plan of this paper is as follows. In Section 2, we consider
two-partners QKD. The material of this section is
not new in itself, but the approach is; moreover, it is a
useful introduction to the following Sections. In Section 3, we
define the N-partners protocols that we consider. Several
eavesdropping scenarios can be imagined on these protocols, and we
give Eve's optimal individual attack in each case. In Section 4,
we introduce a family of M-qubit Bell's inequalities
(Mermin-Klyshko inequalities), and we discuss the link between the
violation of these inequalities and the security of the N-partners
protocol. Section 5 is a conclusion.

\section{QKD involving two partners}

\subsection{The BB84 protocol}

The BB84 protocol of quantum key distribution between two
partners, Alice and Bob, is characterized by the fact that two
complementary bases are used to encode the bits. In the original
version of the BB84 protocol \cite{bb84}, Alice prepares a qubit
into a randomly chosen eigenstate of $\si_x$ or of $\si_y$, and
sends it to Bob. Since we want to discuss Bell's inequalities, we
consider the following preparation method \cite{bbm}: Alice has
an EPR source that produces a maximally entangled state, say
$\ket{\Phi^+_z}=\frac{1}{\sqrt{2}}(\ket{00}+ \ket{11})$, where
$\ket{0}$ and $\ket{1}$ are the eigenstates of $\si_z$. On her
side, Alice measures randomly $\si_x$ or $\si_y$ on one qubit. The
moment at which Alice performs her measurement is irrelevant; in
particular, she can measure her qubit immediately after it leaves
the source. This way, Alice's measurement acts as a preparation of
the second qubit, which goes to Bob through a quantum channel.
Bob also measures either $\si_x$ or $\si_y$. If he measures the
same observable as Alice, his result is perfectly correlated to
hers, since $\moy{\si_x\otimes\si_x}_{\Phi^+}=
-\moy{\si_y\otimes\si_y}_{\Phi^+}=1 $; if he measures the other
observable, he has no information on Alice's result, since
$\moy{\si_x\otimes\si_y}_{\Phi^+}=
\moy{\si_y\otimes\si_x}_{\Phi^+}=0$. At the end of the
transmission, for each qubit Alice and Bob reveal publicly the
measurement that they performed (but of course not its result).
They simply discard those cases where they have measured
different observables, and they end up with two identical lists of
random bits. Bob's information on Alice's bits is measured by the
{\em mutual information} $I(A:B)$, defined as
$H(A)+H(B)-H(AB)=H(A)-H(A|B)$, where $H(\{p_i\})=-\sum_i
p_i\mbox{log}_2p_i$ is the Shannon entropy. Therefore, in the
absence of eavesdropping, $H(A|B)=0$ since knowing the list of B
is equivalent to knowing the list of A; and since Alice is
supposed to choose her measurement randomly, $H(A)=1$; whence
$I(A:B)=1$, as it should.

\subsection{Security and mutual information}

The security of a key-distribution protocol based on quantum
mechanics comes from the no-cloning theorem. Suppose that Eve
tries to eavesdrop on the quantum channel linking Alice and Bob:
she cannot get information on the state that is sent on the
channel without introducing perturbations, that should reveal her
presence to the authorized partners. If A and B observe the
presence of the spy, in most cases they can still perform some
operations that ultimately lead them to share a secret key. More
precisely \cite{csi} A and B can run a one-way protocol known as {\em privacy
amplification} if and only if \be
I(A:B)\,>\,\min[I(A:E),I(B:E)]\,.\label{condsec}\ee This is the
condition that we ask for {\em security}. In fact, it has been
shown that this condition is not strictly necessary: even if it
does not hold, there exist a protocol allowing the extraction of
a secret key \cite{wolf}. But this protocol, called {\em advantage
distillation}, is a two-way protocol, much
less efficient than one-way privacy amplification.

The natural problem is now: for a given error rate that is
introduced on Bob's information, that is, for a given value of
$I(A:B)$, find the attack of Eve that optimizes her information
on Alice's key, $I(A:E)$. The answer is not known in all
generality; but it is, if we restrict the analysis to {\em
individual attacks} \cite{fuchs}. This means that Eve acts
separately on each qubit that is sent on the quantum channel,
i.e., she does not perform coherent measurements of subsequent
qubits. To date, it is not known whether a more general attack
would be more efficient, only bounds are known \cite{shor} --- as
for the experimental state-of-the-art, even the implementation of
individual attacks would be a great challenge.

\begin{figure}
\begin{center}
\epsfbox{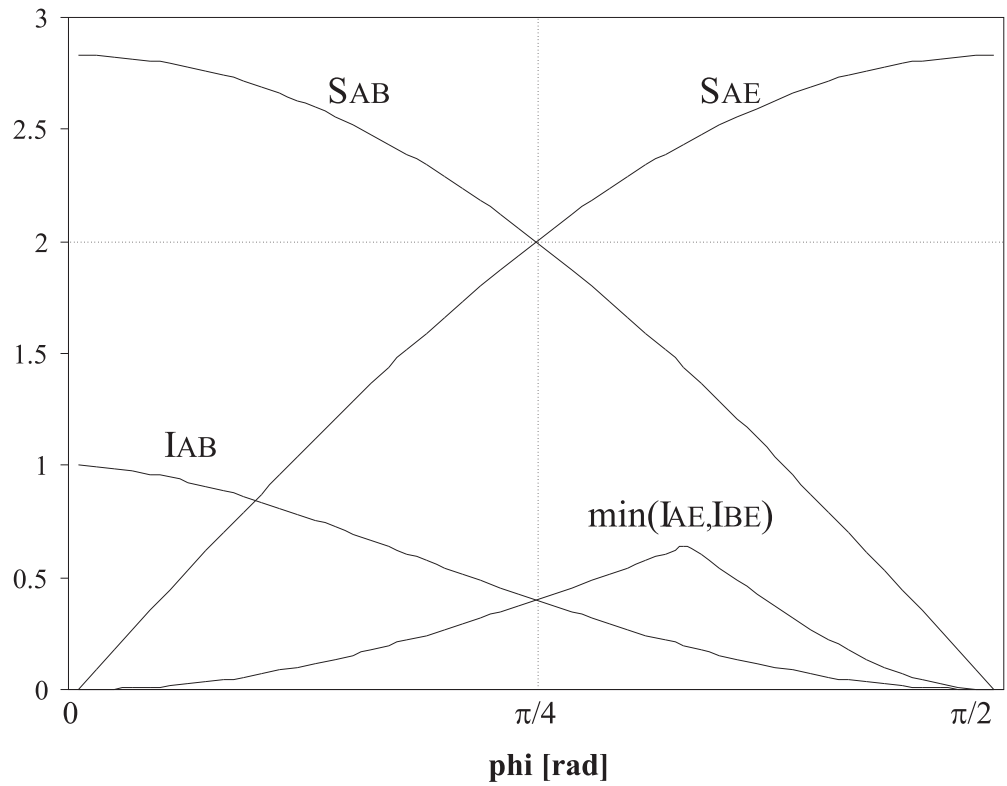}
\caption{Bell parameter $S$ and mutual information $I$ for the two-partners QKD protocol BB84.
The security criterion (\ref{condsec}) is satisfied if and only if
$S_{AB}>2$.} \label{figqkd}
\end{center}
\end{figure}

It has been shown \cite{niu} that Eve can perform the optimal
individual attack having a single qubit as resource, by
implementing the following unitary transformation affecting her
and Bob's qubits (by convention, we supposed that Eve prepares
her qubit in the state $\ket{0}$): \be
\begin{array}{lll}
U_{BE}\ket{00}&=&\ket{00}\\
U_{BE}\ket{10}&=&\cos\phi\ket{10}\,+\,\sin\phi\ket{01}
\end{array}
\label{isometry} \ee where $\ket{00}$ etc. are shorthand for
$\ket{0}_B\otimes\ket{0}_E$ etc.; $\phi\in [0,\frac{\pi}{2}]$
characterizes the strength of Eve's attack. Note that the roles
of B and E are symmetric under the exchange of $\phi$ with
$\frac{\pi}{2}-\phi$. Due to eavesdropping, the three-qubit state
of A, B and E reads
\be\ket{\Psi_{21}(\phi)}\,=\,\frac{1}{\sqrt{2}}
(\ket{000}+\cos\phi \ket{110}+\sin\phi
\ket{101})\,.\label{stateqkd} \ee (the labeling means that we
consider a 2-partners protocol in which 1 partner is spied by
Eve). By tracing out one of the qubits, we obtain the density
matrices $\rho_{AB}$, $\rho_{AE}$ and $\rho_{BE}$ that describe
the statistics of each pair. Let then $M,N\in\{A,B,E\}$, $M\neq
N$, be two of the three partners. We want to calculate $I(M:N)$.
In general, we must consider two statistics: the statistics $p_x$
obtained when $M$ and $N$ measure $\si_x$, and the statistics
$p_y$ obtained when $M$ and $N$ measure $\si_y$; and
$I(M:N)=1-\demi [H_x(M|N)+H_y(M|N)]$. Here, after calculation one
finds that both statistics are the same, whatever the pair.
Moreover, $p(M=0,N=0)=p(M=1,N=1)$ and $p(M=0,N=1)=p(M=1,N=0)$.
Writing $D_{MN}=p(M=0,N=1)+p(M=1,N=0)$ the probability that the
bit of $M$ is different from the bit of $N$, we find finally \ba
I(M:N)&=&1-H(\{D_{MN},1-D_{MN}\})\ea with \be
\begin{array}{ccc}
D_{AB}\,=\,\frac{1-\cos\phi}{2}\,,&
D_{AE}\,=\,\frac{1-\sin\phi}{2}\,,& D_{BE}\,=\,\frac{1-\sin 2
\phi}{2}\,.\end{array} \label{errors}\ee In figure \ref{figqkd},
we plotted $I(A:B)$ versus $\min[I(A:E),I(B:E)]$: we see that the
condition for security (\ref{condsec}) is fulfilled if and only if
$\phi\leq\frac{\pi}{4}$.

\subsection{Violation of a Bell's inequality}

Having the three density matrices $\rho_{AB}$, $\rho_{AE}$ and
$\rho_{BE}$ derived from the three-qubit state (\ref{stateqkd}),
we can also investigate whether one or more pairs violate a
Bell's inequality for a given value of $\phi$. Given a set of four
unit vectors $\underline{a}=\{\vec{a}_1,\vec{a}\,'_1,
\vec{a}_2,\vec{a}\,'_2\}$, we build the two-qubit Bell operator
\ba {\cal{B}}_2(\underline{a})&=&
\left(\sigma_{a_1}+\si_{a'_1}\right)\otimes\si_{a_2} +
\left(\sigma_{a_1}-\si_{a'_1}\right)\otimes\si_{a_2'}
\label{eqchsh}\ea with $\sigma_a=\vec{a}\cdot\vec{\si}$. The CHSH
inequality \cite{chsh} reads
$S_2=\max_{\underline{a}}\Tr(\rho\,{\cal{B}}_2(\underline{a}))
\leq 2$, while the maximal value allowed by QM is $S_2=2\sqrt{2}$
\cite{cirelson}. The calculation of $S_2$ using the Horodecki
criterion \cite{horo} can be carried out explicitly for the three
pairs, and we find \be
\begin{array}{ccc}
S_{AB}\,=\,2\sqrt{2}\,\cos\phi\;, &
S_{AE}\,=\,2\sqrt{2}\,\sin\phi\;,& S_{BE}\,=\,\sqrt{2}\,\sin 2
\phi\;.
\end{array}
\ee Therefore the pair A-B violate the inequality if and only if
the pair A-E does not violate it, and the curves cross at
$\phi=\frac{\pi}{4}$, exactly were the security condition ceases
to be fulfilled (fig. \ref{figqkd}). As for the pair B-E, it never
violates the inequality. The fact that the curves $S_{AB}$ and
$S_{AE}$ cross at $\phi=\frac{\pi}{4}$ is an immediate
consequence of the symmetry of the attack (\ref{isometry}); but
what is interesting is that they cross precisely for
$S_{AB}=S_{AE}=2$. In other words: simply using the symmetry, we
could have guessed that $I(A:B)>I(A:E)$ if and only if
$S_{AB}>S_{AE}$; but here we found that Eve's optimal attack is
such that A and B can establish a secret key using one-way
privacy amplification iff $S_{AB}>2$, i.e., iff they violate the
CHSH inequality. This coincidence was already stressed in
\cite{bruno,fuchs}.

It has recently been shown that the analysis of the BB84 protocol holds
unchanged even if we suppose that Eve controls the source
\cite{hitoshi}. Also, in the
case of the six-state protocol for QKD \cite{sixstate}, the violation
of CHSH is still a sufficient, but no longer a necessary condition.
In conclusion, a {\em tour d'horizon} of the two-partners QKD protocols
with qubits shows that the violation of the CHSH inequality is a
sufficient condition for security. In Section 4, we generalize
this statement to protocols in which the key is distributed among
more than two partners. The next section is devoted to the
definition of these protocols.

\section{QKD involving N partners: definition, and eavesdropping}

\subsection{The N-partners secret-sharing protocol}

The QKD protocol can be generalized to more than two partners in
several ways. For instance, one may think of a protocol in which
Alice sends information to Bob and Charlie so that she can choose
{\em a posteriori} with whom a secret key will be established.
Here we consider another family of protocols, based on the
following idea: Alice sends information to her $N-1$ partners
$B_1, ...,B_{N-1}$ in such a way that {\em all} of them must
cooperate in order to retrieve the secret key, and any smaller
subset of Bobs has no information on the key. More formally, this
means that the bipartite mutual information $I(A:B_1,...,B_k)$
must be 0 for $k<N-1$, and 1 for $k=N-1$. Such protocols exist,
and are called {\em secret-sharing protocols} \cite{qss}.

Now, take $k+1$ of the partners and divide the partners into three
non-empty groups $\cal{A}$, $\cal{B}$ and $\cal{C}$. In general,
it holds $I({\cal{A}}:{\cal{B}}{\cal{C}})
=I({\cal{A}}{\cal{B}}:{\cal{C}})+H({\cal{B}}|{\cal{C}})
-H({\cal{B}}|{\cal{A}})$. But in the protocols that we are
considering, if we know only ${\cal{A}}$, we have no information
on ${\cal{B}}$, since we lack the information of ${\cal{C}}$;
thus $H({\cal{B}}|{\cal{A}})=H({\cal{B}})$. Similarly,
$H({\cal{B}}|{\cal{C}})=H({\cal{B}})$. Consequently for
secret-sharing protocols we have $I({\cal{A}}:{\cal{B}}{\cal{C}})
=I({\cal{A}}{\cal{B}}:{\cal{C}})=I(A:B_1,...,B_{k})$.

The quantum version of a secret sharing protocol involving N
partners (NQSS) goes as follows: Alice's source produces the
N-qubit GHZ state $\ket{\mbox{GHZ}_N}
=\frac{1}{\sqrt{2}}(\ket{0^N}_z+\ket{1^N}_z)$, with
$\ket{0^N}_z=\ket{0}_z\otimes...\otimes \ket{0}_z$, and
$\ket{1^N}_z=\ket{1}_z\otimes...\otimes \ket{1}_z$. Alice
measures $\si_x$ or $\si_y$ on one of the qubits, and sends the
other qubits to her partners $B_1,...,B_{N-1}$. Each Bob also
measures either $\si_x$ or $\si_y$. At the end of the
transmission, all partners communicate publicly their
measurements. Each time that an even number of partners have
measured $\si_y$, the results exhibit the desired correlation. In
fact, consider a measurement where all partners measured $\si_x$:
each partner has one bit $s_{A}$, $s_{B_1}$, ..., $s_{B_{N-1}}$,
where $s=\pm 1$. Since $\moy{\si_x\otimes
...\otimes\si_x}_{GHZ}=1$, these N bits must satisfy
$s_{A}s_{B_1}...s_{B_{N-1}}=1$. Consequently, if all the Bobs
cooperate, they know Alice's bit $s_{A}=s_{B_1}... s_{B_{N-1}}$;
and if one or more of the Bobs refuses to cooperate, then the
other Bobs have strictly no information on what Alice has sent.

Before studying Eve's optimal individual attacks on NQSS, we
introduce the useful notations \[
\begin{array}{lcl}
\ket{0^N}_z\,=\,\ket{0}_z\otimes...\otimes
\ket{0}_z&,&\ket{1^N}_z\,=\,\ket{1}_z\otimes...\otimes \ket{1}_z\\
\ket{0^N}_x\,=\,
\frac{1}{\sqrt{2}}(\ket{0^N}_z+\ket{1^N}_z)&,&\ket{1^N}_x\,=\,
\frac{1}{\sqrt{2}}(\ket{0^N}_z-\ket{1^N}_z)\\
\ket{0^N}_y\,=\,
\frac{1}{\sqrt{2}}(\ket{0^N}_z+i\ket{1^N}_z)&,&\ket{1^N}_y\,=\,
\frac{1}{\sqrt{2}}(\ket{0^N}_z-i\ket{1^N}_z)\,.\end{array} \]
These notations may seem misleading, since $\ket{0^N}_x$,
$\ket{1^N}_x$, $\ket{0^N}_y$ and $\ket{1^N}_y$ are not product
states like $\ket{0^N}_z$ and $\ket{1^N}_z$, but GHZ states. It
will become evident in the following why such notations are
indeed suited to our analysis.

\subsection{Optimal eavesdropping for 3QSS}

For clarity, we discuss in all detail Eve's attacks in the case
$N=3$, that is, on the quantum secret sharing protocol proposed in \cite{qss}.
Alice's authorized partners are called Bob and Charlie. Two
scenarios for eavesdropping can be imagined.

{\em Scenario 1.} An external Eve tries to eavesdrop on both
channels A-B and A-C, in order to gain as much information as
possible on Alice's message. For the analysis of this scenario,
it is convenient to suppose that Alice measures immediately
$\si_x$ or $\si_y$ on her qubit. This way, she prepares the
two-qubit state that is sent to Bob and Charlie, according to
Table \ref{tableqss1}.
\begin{table}
  \centering
  \begin{tabular}{|c c c|}
    \hline Measure of A & Result &State of BC
    \\\hline
     $\si_x$ & $+\,\equiv\,0$ &$\ket{0^2}_x$\\
      & $-\,\equiv\,1$ &$\ket{1^2}_x$\\
     $\si_y$ & $+\,\equiv\,0$ &$\ket{0^2}_y$\\
      & $-\,\equiv\,1$ &$\ket{1^2}_y$ \\ \hline
  \end{tabular}
  \caption{Preparation of the state of BC by measurements of A.}\label{tableqss1}
\end{table}
If we forget the difference in the {\em physical realization} of
the flying bit and stick to the information content of what is
being transmitted, this eavesdropping scenario is identical to
the eavesdropping on the BB84 protocol. Therefore we know an
individual attack that maximizes $I(A:E)$ for a given $I(A:BC)$:
it is given by (\ref{isometry}), replacing $\ket{0}$ with
$\ket{0^2}_z$ on Bob's side. Note that this attack is "coherent",
in the sense that Eve attacks coherently the two qubits flying to
B and C; but is nevertheless an "individual" attack, since each
pair of qubits is attacked separately form the other pairs. This
concludes the study of scenario 1.

{\em Scenario 2.} Bob does not want to cooperate with Charlie in
order to retrieve Alice's message. Consequently, he collaborates
with an Eve that tries to eavesdrop on the line A-C. In this
scenario, two triples come into play: A-B-C and A-B-E, and the
meaningful information measures are $I(A:BC)$ and $I(A:BE)$. Just
as in the analysis of scenario 1, it is useful to recast the
protocol in the following form: by measuring $\si_x$ or $\si_y$,
A and B prepare the state of the qubit that is sent to C. The
preparation is given in Table \ref{tableqss2}.
\begin{table}
  \centering
  \begin{tabular}{|c c c|c c c|}
    \hline
    Meas. of A and B & Result &State of C &
    Meas. of A and B & Result&State of C\\\hline
     $\si_x\otimes\si_x$ & $++$ &$\ket{0}_x$ & $\si_y\otimes\si_x$ & $++$ &$\ket{0}_y$\\
     & $+-$ &$\ket{1}_x$ & & $+-$ &$\ket{1}_y$\\
     & $-+$ &$\ket{1}_x$ & & $-+$ &$\ket{1}_y$\\
     & $--$ &$\ket{0}_x$ & & $--$ &$\ket{0}_y$\\
     $\si_x\otimes\si_y$ & $++$ &$\ket{0}_y$ &  $\si_y\otimes\si_y$ & $++$ &$\ket{1}_x$\\
     & $+-$ &$\ket{1}_y$ & & $+-$ &$\ket{0}_x$\\
     & $-+$ &$\ket{1}_y$ & & $-+$ &$\ket{0}_x$\\
     & $--$ &$\ket{0}_y$ & & $--$ &$\ket{1}_x$ \\ \hline
  \end{tabular}
  \caption{Preparation of the state of C by measurements of A and B.}\label{tableqss2}
\end{table}
The bits flying on the channel A-C are exactly in the same
physical state as in a BB84 protocol. The conclusion here is not
as straightforward as for scenario 1 however, because the
individual attack (\ref{isometry}) optimizes for $I(AB:E)$ with
respect to $I(AB:C)$, while we need the attack that optimizes
$I(A:BE)$ with respect to $I(A:BC)$. But $I(AB:C)=I(A:BC)$ and
$I(AB:E)=I(A:BE)$ hold even in the presence of the eavesdropper:
in fact, B and C are not correlated before the eavesdropping, and E
is not correlated with A and B; therefore
$H(B|A)=H(B|C)=H(B|E)=H(B)$.

\subsection{Optimal eavesdropping for NQSS}

The same argument can be worked out for any eavesdropping
scenario on NQSS, for arbitrary $N$. One the one side, as
discussed, even under eavesdropping, it holds that
$I(AB_1...B_{N-n-1}:B_{N-n}...B_{N-1})\,=\, I(A:B_1...B_{N-1})$.
On the other side, conditioned to the measurements of $\si_x$ and
$\si_y$ by $N-n$ partners, the $n$ other partners can only
receive one of the four states $\ket{0^n}_x$, $\ket{1^n}_x$,
$\ket{0^n}_y$, $\ket{1^n}_y$. In fact, take as an example the
case where all the $N-n$ partners measure $\si_x$. The N-qubit
GHZ state $\ket{\mbox{GHZ}_N}$ can be rewritten as (we neglect
normalization) \ban \ket{0^N}_z+\ket{1^N}_z&\simeq&\big(\ket{0}_x+
\ket{1}_x\big)^{\otimes(N-n)} \ket{0^n}_z+
\big(\ket{0}_x-\ket{1}_x\big)^{\otimes(N-n)}
\ket{1^n}_z\,=\\
&=& (\mbox{even number of $1_x$'s})\otimes\ket{0^n}_x+ (\mbox{odd
number of $1_x$'s})\otimes\ket{1^n}_x\,.\ean Then, conditioned on
the result of the measurement of $\si_x$ on the first $N-n$
qubits, either $\ket{0^n}_x$ or $\ket{1^n}_x$ is sent to the
remaining $n$ partners. The case where some of the $N-n$ partners
measure $\si_y$ is analogous; the states that are prepared are
$\ket{0^n}_x$ or $\ket{1^n}_x$ if an even number of partners
measure $\si_y$, $\ket{0^n}_y$ or $\ket{1^n}_y$ otherwise.\\

In conclusion, we have shown that, for all possible eavesdropping
scenarios on NQSS protocols, Eve can perform the optimal
individual attack having a single qubit as resource. The
interaction that describes the optimal individual attack on $n$
channels is \ba
  \begin{array}{lcl}
    \ket{0^n}_z\otimes\ket{0}_E & \longrightarrow & \ket{0^n}_z\otimes\ket{0}_E \\
    \ket{1^n}_z\otimes\ket{0}_E & \longrightarrow &
    \cos\phi\ket{1^n}_z\otimes\ket{0}_E+ \sin\phi\ket{0^n}_z\otimes\ket{1}_E
  \end{array}
\label{optimaln}\ea where $\phi\in[0,\frac{\pi}{2}]$ measures the
strength of the interaction. Of course, this interaction is
presumably more complicated when she eavesdrops on several
channels (like in scenario 1 for 3QSS), since she must have her
qubit interacting coherently with all flying qubits. For the
attack (\ref{optimaln}), the mutual information for the
authorized and for the unauthorized partners, $I_a$ and $I_u$
respectively, can be calculated explicitly; it is not astonishing
that the result is \ba I_a &\equiv&
I(A:B_1,...,B_{N-1})\,=\,I_{N=2}(A:B)\,=\,1-H(\{D_{AB},
1-D_{AB}\})\\
I_u &\equiv& I(A:B_1,...,B_{N-n-1},E)\,=\,I_{N=2}(A:E)\,=\,
1-H(\{D_{AE}, 1-D_{AE}\}) \ea with $D_{AB}$ and $D_{AE}$ given by
(\ref{errors}). So again by symmetry $I_a>I_u$ if and only if
$\phi<\frac{\pi}{4}$. To our knowledge, privacy amplification has
not been studied in secret-sharing protocols; in particular, it
is not clear if there is still a huge difference in
efficiency between "one-way" and "two-way" protocols. It seems
however obvious that the set of partners having the highest mutual
information can run some protocol to extract a secret key.

\section{Violations of Bell's inequalities}

\subsection{Multiqubit Bell's inequalities}

In the case $N=2$, we saw that $I_a>I_u$ if and only if the
authorized partners violate the CHSH inequality, that is if
$S_a>2>S_u$. We extend this result to all NQSS
protocols.

The number of inequivalent BI grows rapidly with
$M$, the number of qubits. We restrict to the family of inequalities obtained when only
two measurement are performed on each qubit, which are the
natural generalization of the CHSH inequality. Even with this
restriction, the number of possible inequalities grows as
$2^{2^M}$; but recently, the inequalities in this family have been completely
classified by Werner and Wolf \cite{ww}. In particular, these
authors have shown that in this family one can find some
inequalities that are "optimal" under several respects. These
optimal inequalities are nothing but the so-called Mermin-Klyshko
inequalities (MKI) proposed some years ago \cite{mk,helle}. These
are the $M$-qubit BI that we are going to consider in this work.

Let $\underline{a}$ be a set of $2M$ unit vectors. The Bell
operator that enters the MKI for $M$ qubits is defined recursively
as \ba {\cal{B}}_M(\underline{a})\,\equiv\,
{\cal{B}}_M&=&\demi(\si_{a_M}+\si_{a_M'})\otimes
{\cal{B}}_{M-1}\,+\, \demi(\si_{a_M}-\si_{a_M'})\otimes
{\cal{B}}_{M-1}' \label{recurrence}\ea where ${\cal{B}}_m'$ is
obtained from ${\cal{B}}_m$ by exchanging all the $\vec{a}_k$ and
$\vec{a}\,'_k$. The maximal value for product states is
$\moy{{\cal{B}}_M}=2$; quantum correlations allow
$\moy{{\cal{B}}_M}>2$, up to $\moy{{\cal{B}}_M}=
2^{\frac{M+1}{2}}$, obtained for M-qubit GHZ states. It is
important to stress another property of MKIs \cite{helle,wer}.
Let $\rho$ be a M-qubit state, and suppose that you can find a
decomposition $\rho=\sum_i p_i\rho_i$ such that in all $\rho_i$
at most $m<M$ qubits are entangled (not necessarily the same ones
in each $\rho_i$): then $\moy{{\cal{B}}_M}_{\rho}\leq
2^{\frac{m+1}{2}}$. In other words, if
$2^{\frac{m}{2}}<\moy{{\cal{B}}_M}_{\rho}\leq 2^{\frac{m+1}{2}}$,
the inequality for product states is violated, but this violation
is weak, in the sense that it can be achieved with $m$-qubit
entanglement. Now, for MQSS
to work, $\rho$ must be "close" to $\ket{\mbox{GHZ}_M}$,
that is, must exhibit "strong" M-qubit entanglement. Thus in all
that follows we shall say that a M-qubit state $\rho$ {\em
violates the MKI} if the violation is higher than the one that
could be achieved with M-1 qubits, that is, if
$\moy{{\cal{B}}_M}_{\rho}>2^{\frac{M}{2}}$.

\subsection{Violation of MKI in NQSS}

We consider the state that is generated in an eavesdropping
scenario on NQSS. As usual, Alice is the sender. Some of the
receivers would like to retrieve Alice's message without the
other $n$ partners to know it; they ask then Eve to spy on those
lines. We call Bobs $B_1,...,B_{N-n-1}$ the partners that
collaborate with Eve, and Charlies $C_1,...,C_n$ those that are
spied. In the quantum protocol, each partner has a qubit, so we
consider a system of $N+1$ qubits. We write the Hilbert space as
${\cal{H}}_{AB}\otimes{\cal{H}}_C \otimes {\cal{H}}_E $. We have
demonstrated in the previous section that under Eve's optimal
attack the state shared by the $N+1$ partners becomes (we drop
the subscript $z$) \be \ket{\Psi_{Nn}}\,=\,\frac{1}{\sqrt{2}}\,
\big( \ket{0^{N-n}}\ket{0^n}\ket{0}\,+\, \cos\phi\,
\ket{1^{N-n}}\ket{1^n}\ket{0} \,+\, \sin\phi\,
\ket{1^{N-n}}\ket{0^n}\ket{1}\big)\,.\label{psinn} \ee Let
$\rho_{ABC}=\rho_a$ and $\rho_{ABE}=\rho_u$ be the density
matrices of the authorized and of the unauthorized partners that
are derived from $\ket{\Psi_{Nn}}$. We have $S_a\,=\,
\max_{\underline{a}} \mbox{Tr}({\cal{B}}_N(\underline{a})
\rho_a)$ and $S_u\,=\, \max_{\underline{a}}
\mbox{Tr}({\cal{B}}_{N-n+1}(\underline{a}) \rho_u)$. In the
absence of a criterion like Horodeckis' \cite{horo}, it is
difficult to perform the optimization that gives $S$, even for
the particular state that we consider. We found an explicit
result when $N$ and $n$ have different parity, and relied on
numerical optimization for the other cases. These results are
given in Appendix A. Within these warnings, we can safely state
that the following holds: in the NQSS protocol, whatever the
number $n$ of honest partners that are eavesdropped by Eve:
\ba I_a>I_b &\mbox{if and only if}&
S_a\,
>\,2^{\frac{N}{2}}\,; \label{secnqss}\ea and in this case $S_u<2^{\frac{N-n+1}{2}}$.
This is the exact
analog of the result obtained for $N=2$: in case of optimal
attack by Eve, the authorized partners have a higher information
than the unauthorized ones if and only if they violate the MKI.
In other words, to within the warnings above, we have proved the
Conjecture put forward in a previous work \cite{scaprl}.

\subsection{MKI and the structure of the Hilbert space}

The main feature of the link between optimal eavesdropping and
the violation of MKIs is that the authorized partners violate the
inequality {\em if and only if} the unauthorized partners don't.
Of course, it is trivial to loosen this link: non-optimal attacks
can easily be found in which neither set of partners violate an
inequality. Thus so far we have met only states of qubits
characterized by the following property: if a set $\cal{M}$ of
$M$ qubits violate a MKI, then all other sets of qubits having an
overlap with $\cal{M}$ do not violate a MKI. A natural question
is: is this property true for {\em all} possible states of
qubits? If the answer were positive, then for any given state the
violation of MKIs would define a unique partition of the set of
qubits, into subsets of "strongly entangled" qubits. This would
provide an astonishing link between the violation of MK
inequalities and the structure of the Hilbert space.

However, the answer to this question turns out to be {\em
negative}. The simplest counterexample is provided by a system of
four qubits A,B,C and D, where it is possible that both triples
(A,B,C) and (B,C,D) violate the Mermin's inequality. For example,
for $\alpha\approx 0.955$, the state $\cos\alpha
(\ket{0011}+\ket{1100}+ i\ket{0101}+i\ket{1010})/2+\sin\alpha
(i\ket{1001}+ \ket{1111})/\sqrt{2}$ gives $S_{ABC}= S_{BCD}= 3$,
obviously higher than $2\sqrt{2}$ which is the bound for a
three-qubit violation. We have found numerically several more
examples in which a violation of MKIs by two overlapping sets is
allowed; these results are listed in Appendix B. Interestingly,
there are also some cases in which a double violation is {\em
not} possible: thus there is indeed a link between the violation
of MKIs and the structure of the Hilbert space, although this link
may be difficult to unravel. Possibly a stronger link could be
found by using more general inequalities.

In the meantime, the link between violation of MKI and security is
{\em strengthened} by these remarks. In fact, even though there
exist states in the Hilbert space that would allow double
violations of MKI, these states can never be produced in any
eavesdropping scenario \cite{scaprl}.

\section{Conclusion}

We have demonstrated a link between the security of some quantum
key distribution protocols and the violation of some Bell's
inequalities. Precisely: in a secret-sharing protocol, the
authorized partners have a higher mutual information than the
unauthorized ones if and only if they violate a Mermin-Klyshko
inequality. Whether this result is valid for other protocols, or
for other inequalities, is an open question worth investigating.

All the protocols described in this paper can be implemented using
qubits. It is a current field of investigation whether higher
security can be achieved using higher-dimensional quantum systems
\cite{helle2}. Now, for such systems, no satisfactory Bell inequality
has been found yet; the link with cryptography may provide a
pathway to some new advances in this direction.

\section*{Acknowledgements}

We acknowledge partial financial support from the Swiss FNRS and the
Swiss OFES within the European project EQUIP (IST-1999-11053).

\section*{Appendix A}

We want to calculate $S_a\,=\, \max_{\underline{a}}
\mbox{Tr}(B_N(\underline{a}) \rho_a)$ and $S_u\,=\,
\max_{\underline{a}} \mbox{Tr}(B_{N-n+1}(\underline{a}) \rho_u)$,
where $\rho_a=\rho_{ABC}$ and $\rho_u=\rho_{ABE}$ derived from the
state $\ket{\Psi_{Nn}}$ defined in (\ref{psinn}). We discuss
first the calculation of $S_a$.

Let ${\cal{B}}={\cal{B}}_N(\underline{a})\otimes\one_E$. Then we
have $S_a\,=\,\max_{\underline{a}}
\sandwich{\Psi_{Nn}}{{\cal{B}}}{\Psi_{Nn}}$. Now: \ban
\sandwich{\Psi_{Nn}}{{\cal{B}}}{\Psi_{Nn}} &=&\demi\,
\Big[\sandwich{0^N}{{\cal{B}}_N}{0^N}+
\cos^2\phi\,\sandwich{1^N}{{\cal{B}}_N}{1^N} +
\sin^2\phi\,\sandwich{1^{N-n}0^n}
{{\cal{B}}_N}{1^{N-n}0^n}\,+\\&&+\,
\cos\phi\,(\sandwich{1^N}{{\cal{B}}_N}{0^N}+c.c.)\Big]\,. \ean But
$\sandwich{1^N}{{\cal{B}}_N}{1^N}=(-1)^N
\sandwich{0^N}{{\cal{B}}_N}{0^N}$ and
$\sandwich{1^{N-n}0^n}{{\cal{B}}_N}{1^{N-n}0^n} = (-1)^{N-n}
\sandwich{0^N}{{\cal{B}}_N}{0^N}$, as can be easily verified from
the definition of ${\cal{B}}_N$. Therefore \ba S_a&=&
\max_{\underline{a}}\,\big( f_{Nn}(\phi)\,
B_{00}(\underline{a})\,+\, \cos\phi\, B_{10}(\underline{a})\big)
\label{intermediate} \ea where
$B_{00}=\sandwich{0^N}{{\cal{B}}_N}{0^N}$,
$B_{10}=\mbox{Re}\sandwich{1^N}{{\cal{B}}_N}{0^N}$, and where the
function $f_{Nn}(\phi)$ is positive and depends on the parity of
$N$ and $n$. Before discussing it in detail let's see why the
calculation of $S_a$ is not trivial. We know that there are sets
$\underline{a}$ of unit vectors that saturate the bound
$B_{10}(\underline{a})= 2^{\frac{N+1}{2}}$; but for these we find
$B_{00}(\underline{a})= 0$. Similarly, the sets $\underline{a}$
that saturate the bound $B_{00}(\underline{a})= 2$ give
$B_{10}(\underline{a})=0$. Thus to calculate $S_a$ we cannot
optimize both $B_{00}$ and $B_{10}$: we must know whether it is
better to optimize one of the two and letting the other go to
zero, or if we must find an intermediate value. Numerical
estimates suggest that the first strategy is the good one.
However, by considering all possible values of $f_{Nn}(\phi)$ we
can get some more insight
\begin{itemize}
\item For $N$ odd and $n$ even, $f_{Nn}(\phi)=0$. Consequently the
maximization is immediate: $S_a=2^{\frac{N+1}{2}}\cos\phi$, that
goes below the limit $2^{\frac{N}{2}}$ precisely for
$\phi=\frac{\pi}{4}$.
\item For $N$ even and $n$ odd, $f_{Nn}(\phi)=\cos^2\phi$.
Therefore $S_a\leq\cos\phi\,\max_{\underline{a}}\,\big(
B_{00}(\underline{a})\,+\,B_{10}(\underline{a})\big)$. But since
$N$ is even, $B_{00}(\underline{a})\,+\, B_{10}(\underline{a})=
\sandwich{\mbox{GHZ}}{{\cal{B}}_N(\underline{a})}{\mbox{GHZ}}$,
that can reach $2^{\frac{N+1}{2}}$. Consequently the bound can be
achieved, and we have again $S_a=2^{\frac{N+1}{2}}\cos\phi$.
\item  For both $N$ and $n$ odd, $f_{Nn}(\phi)=\sin^2\phi$; and for both $N$ and
$n$ even, $f_{Nn}(\phi)=1$. For these cases, we did not find any
argument leading to a simple estimate of $S_a$. However, several
numerical estimates strongly suggest that $S_{a}=\max\,\left[
2^{\frac{N+1}{2}}\cos\phi\, ,\, 2\,f_{Nn}(\phi)\right]$. In
particular, the boundary $S=2^{\frac{N}{2}}$ is once again crossed
for $\phi=\frac{\pi}{4}$.
\end{itemize}

The same discussion can be made for $S_u$, replacing $\phi$ by
$\frac{\pi}{2}-\phi$, $N$ by $N'=N-n+1$ and $n$ by $n'=1$.
Therefore $n'$ is always odd. If $N$ and $n$ have different
parities, then $N'$ is even, and we have certainly
$S_u=2^{\frac{N-n+1}{2}}\sin\phi$. If $N$ and $n$ have the same
parity, then $N'$ is odd as $n'$, and we are left with numerical
arguments.

In conclusion, the condition for security (\ref{secnqss}) has
been rigorously demonstrated for $N$ and $n$ of different
parity. Note that this case includes the case where $n=N-1$,
that is the case of an external Eve and no dishonest Bob. For the
cases where $N$ and $n$ have the same parity, we did not find a
conclusive demonstration. However, both numerical arguments
\cite{note0} and formal analogies (the structure of the states is
identical) strongly suggest that (\ref{secnqss}) holds in these
cases too.

\section*{Appendix B}

Consider a set $\cal{K}$ of $k$ qubits, and let $\cal{M}$ and
$\cal{N}$ be two different but overlapping subsets of $\cal{K}$
containing respectively $m$ and $n$ qubits. For definiteness,
take $m\leq n$. For a given $\ket{\Psi}\in{\cal{H}}_{\cal{K}}$,
we write $\rho_{{\cal{N}}}$ and $\rho_{{\cal{M}}}$ the density
matrices for the qubits in the two subsets obtained from
$\ket{\Psi}\bra{\Psi}$ by partial traces. We ask if it is
possible to find a state $\ket{\Psi}\in{\cal{H}}_{\cal{K}}$ such
that \ban S_{{\cal{N}}}=\max_{\underline{a}}
\mbox{Tr}\big(\rho_{{\cal{N}}}\,{\cal{B}}_n(\underline{\mathbf{a}})\big)
\,>\,2^{\frac{n}{2}}&\mbox{and}& S_{{\cal{M}}}=
\max_{\underline{A}} \mbox{Tr}\big(\rho_{{\cal{M}}}\,
{\cal{B}}_m(\underline{\mathbf{A}})\big)\,
>\,2^{\frac{m}{2}}\ean where $\underline{\mathbf{a}}$ and
$\underline{\mathbf{A}}$ are sets of, respectively, $2n$ and $2m$
unit vectors. To tackle this question, we define the observable
\[V_{knm}\,=\,{\cal{B}}_n(\underline{\mathbf{a}})\otimes\one_{k-n}\,+\,
2^{\frac{n-m}{2}}\one_{k-m}\otimes {\cal{B}}_m(
\underline{\mathbf{A}}) \,.\] The computer program maximizes the
highest eigenvalue of $V_{knm}$ over all possible choices of
$\underline{\mathbf{a}}$ and $\underline{\mathbf{A}}$. If the
highest eigenvalue does not exceed $2\times 2^{n/2}$, then it is
impossible to find a state that allows both
$S_{{\cal{N}}}>2^{n/2}$ and $S_{{\cal{M}}}>2^{m/2}$. The results
of the numerical calculations that we performed are listed here
(for clarity, we print in boldface the common qubits):
\begin{itemize}
\item $\max(S_{\mathbf{A}B}+S_{\mathbf{A}C})=4$: double violation
impossible (Theorem 1 in \cite{scaprl}).
\item $\max(S_{\mathbf{A}BC}+S_{\mathbf{A}DE})=4\sqrt{2}$: impossible (Theorem 2 in \cite{scaprl}).
\item $\max(S_{\mathbf{AB}C}+S_{\mathbf{AB}D})=6.0945>4\sqrt{2}$: possible (Theorem 3 in \cite{scaprl};
see main text for a state that gives a double violation).
\item $\max(S_{\mathbf{A}BC}+\sqrt{2}\,S_{\mathbf{A}D})=4\sqrt{2}$: impossible.
\item $\max(S_{\mathbf{AB}C}+\sqrt{2}\,S_{\mathbf{AB}})=6.9282>4\sqrt{2}$: possible,
e.g. for the state $\frac{1}{\sqrt{2}}(\ket{000}+
\cos\alpha\ket{111}+\sin\alpha\ket{110})$, $\alpha\in ]0,
\frac{\pi}{2}[$.
\item $\max(S_{\mathbf{A}BCD}+S_{\mathbf{A}DEF})=8$: impossible.
\item $\max(S_{\mathbf{AB}CD}+S_{\mathbf{AB}EF})=8.612>8$:
possible.
\item $\max(S_{\mathbf{ABC}D}+S_{\mathbf{ABC}E})=8$: impossible.
\item $\max(S_{\mathbf{A}BCD}+2\,S_{\mathbf{A}E})=8$:
impossible.
\item $\max(S_{\mathbf{AB}CD}+2\,S_{\mathbf{AB}})=9.7566>8$:
possible.
\item $\max(S_{\mathbf{A}BCD}+\sqrt{2}\,S_{\mathbf{A}EF})=8$:
impossible.
\item $\max(S_{\mathbf{AB}CD}+\sqrt{2}\,S_{\mathbf{AB}E})=6\sqrt{2}>8$:
possible.
\item $\max(S_{\mathbf{ABC}D}+\sqrt{2}\,S_{\mathbf{ABC}})=9.6566>8$:
possible.
\item $\max(S_{\mathbf{AB}CDE}+S_{\mathbf{AB}FGH})=12.088>8\sqrt{2}$:
possible.
\item $\max(S_{\mathbf{ABC}DE}+S_{\mathbf{ABC}FG})=8\sqrt{2}$:
impossible.
\item $\max(S_{\mathbf{ABCD}E}+S_{\mathbf{ABCD}F})=8\sqrt{2}$:
impossible.
\item $\max(S_{\mathbf{AB}CDE}+2\,S_{\mathbf{AB}F})=12.18>8\sqrt{2}$:
possible.
\end{itemize}
In these examples, double violations appear to be possible when
one set is completely contained into the other one i.e.
${\cal{M}}\subset\cal{N}$, or when $\mbox{Card}
({\cal{M}}\cap{\cal{N}})=2$. Of course, it is difficult to guess
general rules from these observations, since we have explored
only the cases $n,m=2,3,4$ and some cases with $n,m=5$.

\newpage

\begin{thebibliography}{10}
\bibitem{note0} The particular case $N=3$, $n=1$ was discussed in detail in
\cite{scaprl}, theorem 3. Other numerical verifications were made
for $N=4$ and $N=5$.
\end{thebibliography}
\end{document}